\documentclass[aps,preprint,amsmath,amssymb,amsfonts,showpacs]{revtex4}
\usepackage{amsthm}
\usepackage{epsfig}
\usepackage{graphicx}
\usepackage{subfigure}
\usepackage{dcolumn}
\usepackage{bm}

\begin{document}

\title{Polyhedra in spacetime from null vectors}

\author{Yasha Neiman}
\email{yashula@gmail.com}
\affiliation{Institute for Gravitation \& the Cosmos and Physics Department, Penn State, University Park, PA 16802, USA}

\date{\today}

\begin{abstract}
We consider convex spacelike polyhedra oriented in Minkowski space. These are the classical analogues of spinfoam intertwiners. We point out a parametrization of these shapes using null face normals, with no constraints or redundancies. Our construction is dimension-independent. In 3+1d, it provides the spacetime picture behind a well-known property of the loop quantum gravity intertwiner space in spinor form, namely that the closure constraint is always satisfied after some $SL(2,C)$ rotation. As a simple application of our variables, we incorporate them in a 4-simplex action that reproduces the large-spin behavior of the Barrett-Crane vertex amplitude. 
\end{abstract}


\maketitle

\section{Introduction}

In loop quantum gravity (LQG) \cite{Rovelli:2004tv,Thiemann:2007zz} and in spinfoam models \cite{Perez:2012wv}, convex polyhedra are fundamental objects. Specifically, the intertwiners between rotation-group representations that feature in these theories can be viewed as quantum versions of convex polyhedra. This makes the parametrization of such shapes a subject of interest for the LQG community. 

In kinematical LQG, one deals with $SU(2)$ intertwiners, which correspond to 3d polyhedra in a local 3d Euclidean frame \cite{Baez:1999tk,Bianchi:2010gc}. These polyhedra are naturally parametrized in terms of area-normal vectors: each face $i$ is associated with a vector $\vec x_i$, such that its norm equals the face area $A_i$, and its direction is orthogonal to the face. The area normals must satisfy a ``closure constraint'':
\begin{align}
 \sum_i \vec x_i = 0 \ . \label{eq:closure_x}
\end{align}
Minkowski's reconstruction theorem guarantees a one-to-one correspondence between space-spanning sets of vectors $\vec x_i$ that satisfy \eqref{eq:closure_x} and convex polyhedra with a spatial orientation. In LQG, the vectors $\vec x_i$ correspond to $SU(2)$ fluxes. The closure condition \eqref{eq:closure_x} then encodes the Gauss constraint, which also generates spatial rotations of the polyhedron. 

In the EPRL/FK spinfoam \cite{EPRL,FK}, the $SU(2)$ intertwiners get lifted into $SL(2,C)$ and acted on by $SL(2,C)$ (Lorentz) rotations. Geometrically, this endows the polyhedra with an orientation in the local 3+1d Minkowski frame of a spinfoam vertex. The polyhedron's orientation is now correlated with those of the other polyhedra surrounding the vertex, so that together they define a generalized 4-polytope (there are issues with shape-matching on shared faces, which are cleanly resolved only in 4-simplices). In analogy with the spatial case, a polyhedron with spacetime orientation can be parametrized by a set of area-normal \emph{simple bivectors} $B_i$. In addition to closure, these bivectors must also satisfy a cross-simplicity constraint:
\begin{align}
 \sum_i B_i = 0\ ;\quad B_i\wedge B_j = 0 \ .
\end{align}
For a discussion of the associated phase space, see e.g. \cite{Baez:1999tk,Dupuis:2011wy}.

In this paper, we present a different parametrization of convex spacelike polyhedra with spacetime orientation. Instead of bivectors $B_i$, we associate null vectors $\ell_i$ to the polyhedron's faces. This parametrization does not require any constraints between the variables on different faces (except for non-degeneracy). It is unusual in that both the area and the full orientation of each face are functions of the data on \emph{all} the faces. Our construction, like the area-vector and area-bivector constructions above, is dimension-independent. Thus, we parametrize $d$-dimensional convex spacelike polytopes with $(d-1)$-dimensional faces, oriented in a $(d+1)$-dimensional Minkowski spacetime. The parametrization is detailed in section \ref{sec:param}. 

In section \ref{sec:simplex}, we use these variables to construct an action principle for a Lorentzian 4-simplex (or its analogue in different dimensions). Our action principle reproduces the large-spin behavior \cite{Baez:2002rx,Freidel:2002mj,Barrett:2002ur} of the Barrett-Crane spinfoam vertex \cite{Barrett:1997gw,Barrett:1999qw}. In particular, it recovers the Regge action for classical simplicial gravity \cite{Regge:1961px}, up to a possible sign and the existence of additional, degenerate solutions. 

In $d=2,3$ spatial dimensions, our parametrization is not really new. It is secretly contained in the spinor-based description \cite{Borja:2010rc,Livine:2011gp} of LQG intertwiners. There, the face normals from \eqref{eq:closure_x} are constructed as squares of spinors (which have an additional phase degree of freedom in $d=3$). It was noticed that the closure constraint in these variables can always be satisfied by acting on the spinors with an $SL(2,C)$ boost. For details at various stages of the spinor formalism's evolution, see \cite{Conrady:2009px,Freidel:2009nu,Freidel:2010tt,Livine:2013tsa}. There is a direct relation between this construction and ours, which we present in section \ref{sec:discuss}. To our knowledge, the simple spacetime picture presented in this paper is new. Hopefully, it will contribute to the geometric interpretation of the modern spinor and twistor \cite{Freidel:2010bw} variables in LQG.

We work with a mostly-plus metric in Minkowski space. When considering actions, we work in units where $c = \hbar = 8\pi G = 1$.  

\section{The parametrization} \label{sec:param}

Consider a set of $N$ null vectors $\ell_i^\mu$ in the $(d+1)$-dimensional Minkowski space $\mathbb{R}^{d,1}$, where $i=1,2,\dots,N$ and $d\geq 2$. We assume the following conditions on the null vectors $\ell_i^\mu$:
\begin{enumerate}
 \item The $\ell_i^\mu$ span the Minkowski space. This implies in particular that $N\geq d+1$.
 \item The $\ell_i^\mu$ are either all future-pointing or all past-pointing.  
\end{enumerate}
The central observation in this paper is that such sets of null vectors are in one-to-one correspondence with convex $d$-dimensional spacelike polytopes oriented in $\mathbb{R}^{d,1}$. The proof is straightforward. First, consider a set $\{\ell_i^\mu\}$ as above. Let us take the sum of the $\ell_i^\mu$, normalized to unit length:   
\begin{align}
 n^\mu = \frac{\sum_i\ell_i^\mu}{\sqrt{-\sum_{i,j}\ell_i\cdot \ell_j}}\ ;\quad n\cdot n = -1 \ . \label{eq:n}
\end{align}
The unit vector $n^\mu$ is timelike, with the same time orientation as the $\ell_i^\mu$. We now take $n^\mu$ to be the unit normal to our spacelike polytope. In other words, we will construct the polytope in the spacelike hyperplane $\Sigma$ orthogonal to $n^\mu$. To do so, we define the projections of the null vectors $\ell_i^\mu$ into this hyperplane:
\begin{align}
 s_i^\mu = \ell_i^\mu + (\ell_i\cdot n)n^\mu \ . \label{eq:s}
\end{align}
The spacelike vectors $s_i^\mu$ automatically sum to zero. Also, since the $\ell_i^\mu$ span the spacetime, the $s_i^\mu$ must span the hyperplane $\Sigma$. By the Minkowski reconstruction theorem, it follows that the $s_i^\mu$ are the $(d-1)$-area normals of a unique convex $d$-dimensional polytope in $\Sigma$. In this way, the null vectors $\ell_i$ define a $d$-polytope oriented in spacetime.

Conversely, let there be a convex $d$-dimensional spacelike polytope oriented in $\mathbb{R}^{d,1}$. Let $\Sigma$ be the polytope's $d$-dimensional hyperplane. Let $s_i^\mu$ be the area-normal vectors to the polytope's $(d-1)$-faces within $\Sigma$. Finally, let $n^\mu$ be the (future-pointing or past-pointing) timelike unit normal to $\Sigma$. We can then construct the set of null vectors $\ell_i^\mu$ parametrizing the polytope by inverting eq. \eqref{eq:s}:
\begin{align}
 \ell_i^\mu = s_i^\mu + \left|s_i\right| n^\mu \ .
\end{align}

Let us now discuss some basic features of the parametrization. The vectors $\ell_i^\mu$ are associated to the polytope's $(d-1)$-dimensional faces. It is clear from the above construction that they are in fact \emph{null normals} to these faces. Specifically, a future-pointing (past-pointing) vector $\ell_i^\mu$ is the future-outgoing (past-outgoing) null normal to the associated face. Of course, one could also change signs in the construction, so that the $s_i^\mu$ and $\ell_i^\mu$ are ingoing normals. In section \ref{sec:simplex}, both possibilities will be used. Now, the orientation of a spacelike $(d-1)$-plane in  $\mathbb{R}^{d,1}$ is in one-to-one correspondence with the directions of its two null normals. Thus, each $\ell_i^\mu$ carries \emph{partial} information about the orientation of the $i$'th face. The second null normal to the face is a function of \emph{all} the $\ell_i^\mu$. It can be expressed as:
\begin{align}
 \tilde\ell_i^\mu = \ell_i^\mu - 2s_i^\mu = -2(\ell_i\cdot n)n^\mu - \ell_i^\mu \ ,
\end{align}
where we recall that $n^\mu$ is given by \eqref{eq:n}. Similarly, the area $A_i$ of each face is a function of the null normals $\ell_i^\mu$ to all the faces:
\begin{align}
 A_i = \left|s_i\right| = -\ell_i\cdot n \ . \label{eq:A}
\end{align}
Finally, the total area of the faces has the simple expression:
\begin{align}
 \sum_i A_i = \sum_i\left|s_i\right| = \sqrt{\textstyle -\sum_{i,j}\ell_i\cdot\ell_j} \ .
\end{align}

\section{A $(d+1)$-simplex action} \label{sec:simplex}

\subsection{Definition}

As a sample application of the null-normal variables, we will now use them to construct a $(d+1)$-simplex action that reproduces (in the $d=3$ case) the large-spin behavior of the Barrett-Crane spinfoam vertex.

Consider a $(d+1)$-simplex in $\mathbb{R}^{d,1}$. Let the index $a=0,1,\dots,d+1$ run over its $d$-dimensional hyperfaces. These hyperfaces are $d$-simplices, which we take to be spacelike. The $a$'th $d$-simplex shares a common $(d-1)$-face with every other $d$-simplex. The face shared with the $b$'th $d$-simplex will be denoted as $ab$. We can thus parametrize the $a$'th $d$-simplex with a set of $d+1$ future-pointing null normals $\ell_{ab}^\mu$, where $b\neq a$. Note that the vectors $\ell_{ab}^\mu$ and $\ell_{ba}^\mu$ are null normals to the same face.

At the level of degree-of-freedom counting, the shape of a $(d+1)$-simplex is determined by the $(d+1)(d+2)/2$ areas $A_{ab}$ of its $(d-1)$-faces. These areas are directly analogous to the spins that appear in the Barrett-Crane spinfoam. Let us fix a set of values for the $A_{ab}$ and consider the action:
\begin{align}
 S = \sum_{a<b}\left( A_{ab}\ln\left(-\frac{\ell_{ab}\cdot\ell_{ba}}{2A_{ab}^2} \right)
  + \lambda_{ab}(\ell_{ab}\cdot n_a + A_{ab}) + \lambda_{ba}(\ell_{ba}\cdot n_b + A_{ab}) \right) \ . \label{eq:S}
\end{align}
Here, the $\ell_{ab}^\mu$ are null vectors, with no a-priori relation to the geometry of the $(d+1)$-simplex; the relation will emerge dynamically. The $n_a^\mu$ are future-pointing unit timelike vectors. They will emerge as the unit normals to the $d$-simplices, but this is again not fixed a-priori. Finally, the $\lambda_{ab}$ in \eqref{eq:S} are Lagrange multipliers that fix the products $-\ell_{ab}\cdot n_a$ to the corresponding face areas, as in \eqref{eq:A}. One could also introduce Lagrange multipliers to enforce the null and unit nature of $\ell_{ab}^\mu$ and $n_a^\mu$, respectively. Instead, we will simply restrict to variations where: 
\begin{align}
 \delta\ell_{ab}\cdot\ell_{ab} = \delta n_a\cdot n_a = 0 \ . \label{eq:constraints}
\end{align}
In $d\leq 3$, one could make the $\ell_{ab}^\mu$ automatically null by expressing them as products of spinors. For our purposes, vector language will suffice.

\subsection{Stationary point analysis}

In $d=3$, the action \eqref{eq:S} has the same stationary points, and takes the same values there, as the effective large-spin action for the Barrett-Crane vertex. In other dimensions, the behavior is completely analogous. In particular, at non-degenerate stationary points, i.e. ones where the $n_a^\mu$ span $\mathbb{R}^{d,1}$, the 4-simplex geometry is recovered (up to reflections), and the action reduces to the Regge action (up to sign). 

To show this, let us examine the stationary-point equations:
\begin{align}
 0 = \frac{\delta S}{\delta\lambda_{ab}} &= \ell_{ab}\cdot n_a + A_{ab} \label{eq:d_lambda} \\
 \ell_{ab}^\mu \sim \frac{\delta S}{\delta\ell_{ab,\mu}} &= \frac{A_{ab}\ell_{ba}^\mu}{\ell_{ab}\cdot\ell_{ba}} + \lambda_{ab} n_a^\mu \label{eq:d_ell} \\
 n_a^\mu \sim \frac{\delta S}{\delta n_{a,\mu}} &= \sum_{b\neq a} \lambda_{ab}\ell_{ab}^\mu \ . \label{eq:d_n}  
\end{align}
In the last two lines, we took into account the constraint \eqref{eq:constraints} on $\delta\ell_{ab}^\mu$ and $\delta n_a^\mu$. Let us examine the different components of eq. \eqref{eq:d_ell}. The projection into the $(d-1)$-plane orthogonal to $\ell_{ab}^\mu$ and $n_a^\mu$ shows that the vectors $\ell_{ab}^\mu$, $\ell_{ba}^\mu$ and $n_a^\mu$ are coplanar. This leaves the contraction of \eqref{eq:d_ell} with $\ell_{ab}^\mu$, which fixes the value of the Lagrange multiplier $\lambda_{ab}$:
\begin{align}
 \lambda_{ab} = -\frac{A_{ab}}{\ell_{ab}\cdot n_a} = 1 \ , \label{eq:lambda}
\end{align}
where in the last equality we used eq. \eqref{eq:d_lambda}. Plugging this result into \eqref{eq:d_n}, we find that the unit vector $n_a^\mu$ must be the normal to the $d$-simplex defined by the $\ell_{ab}^\mu$'s:
\begin{align}
 n_a^\mu = \frac{\sum_{b\neq a}\ell_{ab}^\mu}{\sqrt{-\sum_{b,c\neq a}\ell_{ab}\cdot \ell_{ac}}} \ . \label{eq:n_a} 
\end{align}

To sum up, the stationary points of the action \eqref{eq:S} have the following properties. For each $a$, the vectors $\ell_{ab}^\mu$ define a $d$-simplex with unit normal $n_a^\mu$ and $(d-1)$-face areas $A_{ab}$. The $d$-simplices automatically agree on the areas of their shared $(d-1)$-faces. Moreover, we've seen that $n_a^\mu$ is coplanar with $\ell_{ab}^\mu$ and $\ell_{ba}^\mu$. Since the same conclusion can be reached for $n_b^\mu$, this implies that $(n_a^\mu,n_b^\mu,\ell_{ab}^\mu,\ell_{ba}^\mu)$ are all coplanar. Now, in the $a$'th $d$-simplex, the plane orthogonal to the $ab$ face is spanned by $n_a^\mu$ and $\ell_{ab}^\mu$. Similarly, the plane orthogonal to the $ba$ face in the $b$'th $d$-simplex is spanned by $n_b^\mu$ and $\ell_{ba}^\mu$. We conclude that the two $d$-simplices agree not only on the area of their shared $(d-1)$-face, but also on the orientation of its $(d-1)$-plane in spacetime. In other words, they agree on the face's area-normal bivector:
\begin{align}
 B_{ab} = n_a\wedge\ell_{ab} = -n_b\wedge\ell_{ba} = -B_{ba}\ ;\quad \left|B_{ab}\right| = \left|B_{ba}\right| = A_{ab} \ . \label{eq:B_symmetry}  
\end{align}
The relative sign is due to the fact that $\ell_{ab}^\mu$ and $\ell_{ba}^\mu$ point along two \emph{different} null directions in the plane orthogonal to the $ab$ face. Otherwise, the scalar product $\ell_{ab}\cdot\ell_{ba}$ would vanish, making the action \eqref{eq:S} divergent. The area bivectors defined in \eqref{eq:B_symmetry} automatically satisfy closure (which follows from \eqref{eq:n_a}) and cross-simplicity:
\begin{align}
 \sum_{b\neq a} B_{ab} = 0\ ;\quad B_{ab}\wedge B_{ac} = 0 \ .
\end{align} 
We conclude that our stationary points are in one-to-one correspondence with the bivector geometries of \cite{Barrett:1997gw} (Hodge-dualized and generalized to arbitrary dimension), minus the non-degeneracy conditions.

Now, to make the connection with the Barrett-Crane vertex more explicit, let us ``integrate out'' the $\lambda_{ab}$ and $\ell_{ab}^\mu$, expressing the action in terms of the $n_a^\mu$. This means imposing eqs. \eqref{eq:d_lambda}-\eqref{eq:d_ell}, but not eq. \eqref{eq:d_n}. The $\lambda_{ab}$ terms in the action then vanish, leaving us with:
\begin{align}
 S = \sum_{a<b} A_{ab}\ln\left(-\frac{\ell_{ab}\cdot\ell_{ba}}{2A_{ab}^2} \right) \ . \label{eq:S_shell}
\end{align}
Each logarithm in \eqref{eq:S_shell} is determined up to sign by the $n_a^\mu$. To see this, consider first the degenerate case $n_a^\mu = n_b^\mu$, i.e. $n_a\cdot n_b = -1$. Then the area-fixing condition \eqref{eq:d_lambda} and the coplanarity of $(\ell_{ab}^\mu,\ell_{ba}^\mu,n_a^\mu)$ force $\ell_{ab}^\mu$ and $\ell_{ba}^\mu$ to take the form:
\begin{align}
 \ell_{ab}^\mu = A_{ab}(n_a^\mu + \hat s_{ab}^\mu)\ ;\quad \ell_{ba}^\mu = A_{ab}(n_a^\mu - \hat s_{ab}^\mu) \ ,
\end{align}
for some spacelike unit vector $\hat s_{ab}^\mu$ orthogonal to $n_a^\mu$. This fixes the argument of the logarithm in \eqref{eq:S_shell} to:
\begin{align}
 -\frac{\ell_{ab}\cdot\ell_{ba}}{2A_{ab}^2} = 1 \ . \label{eq:degen}
\end{align}
Consider now the non-degenerate case, where $n_a^\mu$ and $n_b^\mu$ are linearly independent. $\ell_{ab}^\mu$ and $\ell_{ba}^\mu$ are then forced to point in the two null directions within the 1+1d plane spanned by $(n_a^\mu,n_b^\mu)$. There is a twofold ambiguity here, since we must choose which of $\ell_{ab}^\mu$ and $\ell_{ba}^\mu$ points along which of the null directions. Once the directions of $\ell_{ab}^\mu$ and $\ell_{ba}^\mu$ are chosen, their extents are determined by the area-fixing condition \eqref{eq:d_lambda}. Overall, $\ell_{ab}^\mu$ is given by:
\begin{align}
 \ell_{ab}^\mu = A_{ab}\left(n_a^\mu \mp \epsilon_{ab}\cdot \frac{n_b^\mu + (n_a\cdot n_b)n_a^\mu}{\sqrt{(n_a\cdot n_b)^2 - 1}} \right) \ . \label{eq:ell}
\end{align}
Here, $\epsilon_{ab}$ is a sign factor, defined as $\epsilon_{ab} = +1$ for ``thick wedges'' (figure \ref{fig:orientation}(i,iii)) and $\epsilon_{ab} = -1$ for ``thin wedges'' (figure \ref{fig:orientation}(ii,iv)). With this definition, a minus sign in front of the $\epsilon_{ab}$ in \eqref{eq:ell} yields the configurations in figure \ref{fig:orientation}(i,ii), while a plus sign leads to figure \ref{fig:orientation}(iii,iv). This peculiar decomposition of the overall sign will serve to simplify the result below. The expression for $\ell_{ba}^\mu$ is identical to \eqref{eq:ell}, with $n_a^\mu$ and $n_b^\mu$ interchanged. The argument of the logarithm in \eqref{eq:S_shell} then reads:
\begin{figure}%
\centering%
\includegraphics[scale=0.75]{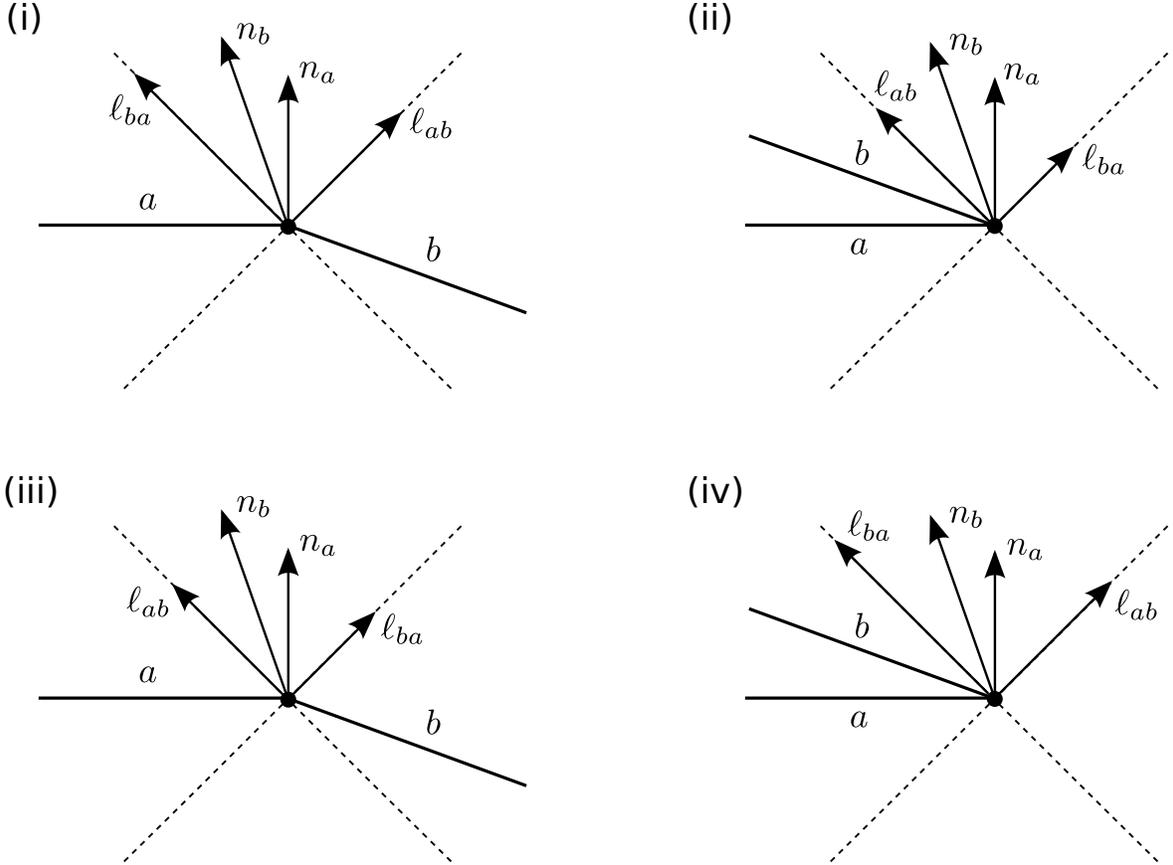} \\
\caption{A $(d-1)$-face in a $(d+1)$-simplex, shared by two $d$-simplices $a$ and $b$. We depict the 1+1d plane orthogonal to the face. The dashed lines are the two null rays in this normal plane. In figures (i) and (iii), both $d$-simplices are ``final'', while in (ii) and (iv), $a$ is initial and $b$ is final. In (i) and (ii), the timelike $d$-simplex normals $(n_a^\mu,n_b^\mu)$ and the null $(d-1)$-face normals $(\ell_{ab}^\mu,\ell_{ba}^\mu)$ correspond to a stationary point of the action \eqref{eq:S} with $S = S_{\text{Regge}}$. Similarly, figures (iii) and (iv) depict a configuration with $S = -S_{\text{Regge}}$.}
\label{fig:orientation} 
\end{figure}%
\begin{align}
 -\frac{\ell_{ab}\cdot\ell_{ba}}{2A_{ab}^2} = -(n_a\cdot n_b) \pm \epsilon_{ab}\sqrt{(n_a\cdot n_b)^2 - 1} \ , \label{eq:log_arg}
\end{align}
for which \eqref{eq:degen} is a special case. Now, notice that the boost angle $\theta(n_a,n_b)$ between $n_a^\mu$ and $n_b^\mu$ is given (up to sign) by:
\begin{align}
 \cosh\theta(n_a,n_b) = -(n_a\cdot n_b) \ .
\end{align}
Plugging this into \eqref{eq:log_arg}, we get:
\begin{align}
 -\frac{\ell_{ab}\cdot\ell_{ba}}{2A_{ab}^2} = \cosh\theta(n_a,n_b) \pm \epsilon_{ab}\left|\sinh\theta(n_a,n_b)\right| 
   = e^{\pm\epsilon_{ab}\left|\theta(n_a,n_b)\right|} \ .
\end{align}
This brings the action to the form:
\begin{align}
 S = \sum_{a<b}\pm \epsilon_{ab} A_{ab}\left|\theta(n_a,n_b)\right| \ , \label{eq:S_BC}
\end{align}
where the sign can be chosen separately for each face $ab$. Eq. \eqref{eq:S_BC} is the effective action for the Lorentzian Barrett-Crane 4-simplex, as studied in \cite{Barrett:2002ur}. At the stationary points, there are two consistent sign choices in \eqref{eq:ell},\eqref{eq:S_BC}. In the first choice, we pick the upper signs in \eqref{eq:ell},\eqref{eq:S_BC} for all the faces, as in figure \ref{fig:orientation}(i-ii). This makes the null normals $\ell_{ab}^\mu$ future-outgoing when the $d$-simplex $a$ is ``final'', and future-ingoing when it is ``initial''. In the second choice, we pick the lower signs in \eqref{eq:ell},\eqref{eq:S_BC} for all the faces, as in figure \ref{fig:orientation}(iii-iv). The $\ell_{ab}^\mu$ are then future-ingoing for final $d$-simplices and vice versa. When the stationary point is non-degenerate, i.e. when the $n_a^\mu$ span the spacetime, the action \eqref{eq:S_BC} reduces to the Regge action, up to sign. For the sign choice corresponding to figure \ref{fig:orientation}(i-ii), we get $S = S_{\text{Regge}}$. For the sign choice corresponding to figure \ref{fig:orientation}(iii-iv), we get $S = -S_{\text{Regge}}$.

\section{Discussion} \label{sec:discuss}

In this paper, we constructed a parametrization for convex spacelike polyhedra (or their dimensional generalizations) oriented in spacetime. The parametrization uses null face normals, which become spacelike area normals once projected into the hyperplane orthogonal to their sum. As a sample exercise with these variables, we incorporated them into a gravitational action for a spacetime simplex. 

As noted in the Introduction, our construction has already appeared in disguise within the LQG literature, in the context of spinor variables. Let us now detail the relation between the two pictures. Throughout this paper, we worked directly in spacetime. In LQG, instead one usually starts with boundary states defined in \emph{space} (actually, a spacelike hypersurface in time gauge). There, one constructs polyhedra in terms of spatial area-normal vectors $\vec x_i$, which satisfy the closure constraint \eqref{eq:closure_x}. In the spinor approach, one expresses the $\vec x_i$ as squares $z_i\bar z_i$ of $SU(2)$ spinors. Now, as discussed in \cite{Conrady:2009px}, if the closure constraint \eqref{eq:closure_x} is not satisfied, one can always recover it by performing an $SL(2,C)$ transformation on the $z_i$. The connection with our picture is as follows. When the $SU(2)$ spinors $z_i$ are reinterpreted as $SL(2,C)$ spinors, their square $z_i\bar z_i$ acquires a new meaning, as a null vector in spacetime. These are precisely our null face normals $\ell_i^\mu$, of which the original $\vec x_i$ are the spatial components! The failure of the $\vec x_i$ to close simply reflects the fact that the $\ell_i^\mu$ are projected into the wrong hyperplane: instead of the polyhedron's hyperplane as determined by the $\ell_i^\mu$ themselves, they are projected into the arbitrary reference hyperplane which was taken as ``space'' in the LQG construction. The $SL(2,C)$ boost described in \cite{Conrady:2009px} reorients the polyhedron into the reference hyperplane. Once this is done, the spatial components of the $\ell_i^\mu$ close.

We conclude with a remark on the time-orientation of the normal vectors in the action \eqref{eq:S}. As in the Barrett-Crane amplitude, we take all the normals to be future-pointing. This makes their scalar products negative, ensuring that the logarithms in \eqref{eq:S} are real. However, in recent papers \cite{Neiman:2013ap,Bodendorfer:2013hla,Neiman:2013lxa}, it has been emphasized by the author that the action of General Relativity has an imaginary part. This imaginary part follows from the $n\pi i/2$ contributions to boost angles that arise when one crosses null directions in a timelike plane \cite{SorkinThesis}. In the present context of a simplex with spacelike faces, these appear as imaginary parts $\pi i$ in the corner angles at ``thin wedges'' (figure \ref{fig:orientation}(ii,iv)). The latter can be incorporated into the action \eqref{eq:S} by changing the time-orientation of $n_a^\mu$ and $\ell_{ab}^\mu$ on initial $d$-simplices $a$ to past-pointing. This means taking all the normals to be outgoing with respect to the $(d+1)$-simplex, rather than taking them all future-pointing. This is of course the necessary choice for the normal that defines the extrinsic curvature in the York-Gibbons-Hawking boundary term \cite{York:1972sj,Gibbons:1976ue} for the continuum action. In the action \eqref{eq:S}, it will result in a negative argument in the logarithm for thin wedges, producing an imaginary part $\pi i$ in the logarithm's result (with the added simplification that the $\epsilon_{ab}$ sign factors in eq. \eqref{eq:ell} become unnecessary). Finally, we note that in the EPRL/FK spinfoam, the large-spin limit of the 4-simplex amplitude automatically ``knows'' about the action's imaginary part: as shown in \cite{Bodendorfer:2013hla}, it can be recovered by sending the Immirzi parameter to $\pm i$ at the end of the stationary-point calculation.

\section*{Acknowledgements}		

I am grateful to Norbert Bodendorfer, Etera Livine and Carlo Rovelli for discussions. This work is supported in part by the NSF grant PHY-1205388 and the Eberly Research Funds of Penn State.


\begin{thebibliography} {99}

\bibitem{Rovelli:2004tv} 
  C.~Rovelli,
  ``Quantum gravity,''
  Cambridge, UK: Univ. Pr. (2004) 455 p

\bibitem{Thiemann:2007zz} 
  T.~Thiemann,
  ``Modern canonical quantum general relativity,''
  Cambridge, UK: Cambridge Univ. Pr. (2007) 819 p
  [gr-qc/0110034].

\bibitem{Perez:2012wv} 
  A.~Perez,
  ``The Spin Foam Approach to Quantum Gravity,''
  Living Rev.\ Rel.\  {\bf 16}, 3 (2013)
  [arXiv:1205.2019 [gr-qc]].

\bibitem{Baez:1999tk} 
  J.~C.~Baez and J.~W.~Barrett,
  ``The Quantum tetrahedron in three-dimensions and four-dimensions,''
  Adv.\ Theor.\ Math.\ Phys.\  {\bf 3}, 815 (1999)
  [gr-qc/9903060].

\bibitem{Bianchi:2010gc} 
  E.~Bianchi, P.~Dona and S.~Speziale,
  ``Polyhedra in loop quantum gravity,''
  Phys.\ Rev.\ D {\bf 83}, 044035 (2011)
  [arXiv:1009.3402 [gr-qc]].

\bibitem{EPRL}
  J.~Engle, E.~Livine, R.~Pereira and C.~Rovelli,
  ``LQG vertex with finite Immirzi parameter,''
  Nucl.\ Phys.\  B {\bf 799}, 136 (2008)
  [arXiv:0711.0146 [gr-qc]].

\bibitem{FK}
  L.~Freidel, K.~Krasnov,
  ``A New Spin Foam Model for 4d Gravity,''
  Class.\ Quant.\ Grav.\  {\bf 25}, 125018 (2008).
  [arXiv:0708.1595 [gr-qc]].

\bibitem{Dupuis:2011wy}
  M.~Dupuis, L.~Freidel, E.~R.~Livine and S.~Speziale,
  ``Holomorphic Lorentzian Simplicity Constraints,''
  J.\ Math.\ Phys.\  {\bf 53} (2012) 032502
  [arXiv:1107.5274 [gr-qc]].

\bibitem{Baez:2002rx} 
  J.~C.~Baez, J.~D.~Christensen and G.~Egan,
  ``Asymptotics of 10j symbols,''
  Class.\ Quant.\ Grav.\  {\bf 19}, 6489 (2002)
  [gr-qc/0208010].

\bibitem{Freidel:2002mj} 
  L.~Freidel and D.~Louapre,
  ``Asymptotics of 6j and 10j symbols,''
  Class.\ Quant.\ Grav.\  {\bf 20}, 1267 (2003)
  [hep-th/0209134].

\bibitem{Barrett:2002ur} 
  J.~WBarrett and C.~M.~Steele,
  ``Asymptotics of relativistic spin networks,''
  Class.\ Quant.\ Grav.\  {\bf 20}, 1341 (2003)
  [gr-qc/0209023].

\bibitem{Barrett:1997gw} 
  J.~W.~Barrett and L.~Crane,
  ``Relativistic spin networks and quantum gravity,''
  J.\ Math.\ Phys.\  {\bf 39}, 3296 (1998)
  [gr-qc/9709028].

\bibitem{Barrett:1999qw} 
  J.~W.~Barrett and L.~Crane,
  ``A Lorentzian signature model for quantum general relativity,''
  Class.\ Quant.\ Grav.\  {\bf 17}, 3101 (2000)
  [gr-qc/9904025].

\bibitem{Regge:1961px} 
  T.~Regge,
  ``General Relativity Without Coordinates,''
  Nuovo Cim.\  {\bf 19}, 558 (1961).

\bibitem{Borja:2010rc} 
  E.~F.~Borja, L.~Freidel, I.~Garay and E.~R.~Livine,
  ``U(N) tools for Loop Quantum Gravity: The Return of the Spinor,''
  Class.\ Quant.\ Grav.\  {\bf 28}, 055005 (2011)
  [arXiv:1010.5451 [gr-qc]].

\bibitem{Livine:2011gp} 
  E.~R.~Livine and J.~Tambornino,
  ``Spinor Representation for Loop Quantum Gravity,''
  J.\ Math.\ Phys.\  {\bf 53}, 012503 (2012)
  [arXiv:1105.3385 [gr-qc]].

\bibitem{Conrady:2009px} 
  F.~Conrady and L.~Freidel,
  ``Quantum geometry from phase space reduction,''
  J.\ Math.\ Phys.\  {\bf 50}, 123510 (2009)
  [arXiv:0902.0351 [gr-qc]].

\bibitem{Freidel:2009nu} 
  L.~Freidel, K.~Krasnov and E.~R.~Livine,
  ``Holomorphic Factorization for a Quantum Tetrahedron,''
  Commun.\ Math.\ Phys.\  {\bf 297}, 45 (2010)
  [arXiv:0905.3627 [hep-th]].

\bibitem{Freidel:2010tt} 
  L.~Freidel and E.~R.~Livine,
  ``U(N) Coherent States for Loop Quantum Gravity,''
  J.\ Math.\ Phys.\  {\bf 52}, 052502 (2011)
  [arXiv:1005.2090 [gr-qc]].

\bibitem{Livine:2013tsa} 
  E.~R.~Livine,
  ``Deformations of Polyhedra and Polygons by the Unitary Group,''
  arXiv:1307.2719 [math-ph].

\bibitem{Freidel:2010bw} 
  L.~Freidel and S.~Speziale,
  ``From twistors to twisted geometries,''
  Phys.\ Rev.\ D {\bf 82}, 084041 (2010)
  [arXiv:1006.0199 [gr-qc]].

\bibitem{Neiman:2013ap} 
  Y.~Neiman,
  ``The imaginary part of the gravity action and black hole entropy,''
  JHEP {\bf 1304}, 071 (2013)
  [arXiv:1301.7041 [gr-qc]].

\bibitem{Bodendorfer:2013hla} 
  N.~Bodendorfer and Y.~Neiman,
  ``Imaginary action, spinfoam asymptotics and the 'transplanckian' regime of loop quantum gravity,''
  arXiv:1303.4752 [gr-qc].

\bibitem{Neiman:2013lxa} 
  Y.~Neiman,
  ``The imaginary part of the gravitational action at asymptotic boundaries and horizons,''
  arXiv:1305.2207 [gr-qc].

\bibitem{SorkinThesis}
  R.~Sorkin,
  ``Development of simplectic methods for the metrical and electromagnetic fields,''
  Ph.D. thesis, California Institute of Technology, 1974. 

\bibitem{York:1972sj} 
  J.~W.~York, Jr.,
  ``Role of conformal three geometry in the dynamics of gravitation,''
  Phys.\ Rev.\ Lett.\  {\bf 28}, 1082 (1972).

\bibitem{Gibbons:1976ue} 
  G.~W.~Gibbons and S.~W.~Hawking,
  ``Action Integrals and Partition Functions in Quantum Gravity,''
  Phys.\ Rev.\ D {\bf 15}, 2752 (1977).

\end{thebibliography}
\end{document}